\documentstyle[prl,aps,epsf]{revtex}
\begin{document}

\newcommand{\nwc}{\newcommand}
\nwc{\ci}[1]{\cite{#1}}

\tolerance=10000

\twocolumn[\hsize\textwidth\columnwidth\hsize
     \csname @twocolumnfalse\endcsname

\title{\bf Spin Chains in an External Magnetic Field.\break
Closure of the Haldane Gap and Effective Field Theories.}
\author{L.Campos Venuti$^{1}$, E.Ercolessi$^{2,3,4}$, G.Morandi$%
^{2,3,4}$, P.Pieri$^{5,6}$ and M.Roncaglia$^{2,3,4}$\\
\begin{center}
{\small {\em $^1$ Institut f\"ur Theoretische Physik 3. Universit\"at 
Stuttgart.}}\\
{\small {\em Pfaffenwaldring 57. D-70550 Stuttgart, Germany.}}\\
{\small {\em $^2$ Physics Dept., University of Bologna, 6/2 v.le B.Pichat, 
I-40127, Bologna, Italy.}}\\
{\small {\em $^3$ INFM, Unit\`a di Bologna, 6/2 v.le B.Pichat, 
I-40127, Bologna, Italy.}}\\
{\small {\em $^4$ INFN, Sezione di Bologna, 6/2 v.le B.Pichat, 
I-40127, Bologna, Italy.}}\\
{\small {\em $^5$ Physics Dept., University of Camerino, 
Camerino, Italy.}}\\
{\small {\em $^6$ INFM, Unit\`a di Camerino, Camerino, Italy.}}\\
{\small {\em E-Mail: MORANDI@BO.INFN.IT.}}\\
\end{center}}
\maketitle
\begin{abstract}
We investigate both numerically and analytically the behaviour of a spin-1
antiferromagnetic (AFM) isotropic Heisenberg chain in an external magnetic
field. Extensive DMRG studies of chains up to $N=80$ sites extend previous
analyses and exhibit the well known phenomenon of the closure of the Haldane
gap at a lower critical field $H_{c_{1}}.$ We obtain an estimate of the gap
below $H_{c_{1}}$. Above the lower critical field, when the correlation
functions exhibit algebraic decay, we obtain the critical exponent as a
function of the net magnetization as well as the magnetization curve up to
the saturation (upper critical) field $H_{c_{2}}.$ We argue that, despite
the fact that the $SO(3)$ symmetry of the model is explicitly broken by the
field, the Haldane phase of the model is still well described by an $SO(3)$
nonlinear $\sigma $-model. A mean-field theory is developed for the latter
and its predictions are compared with those of the numerical analysis and
with the existing literature.
\end{abstract}

\vskip2pc]
\newpage

\section{Introduction.}

Spin chains, ladders and, more generally, low-dimensional quantum spin
systems have become a subject of considerable theoretical and (more recently) 
experimental interest \ci{AMS}. 
This has been due in the recent past to the connection that 2D AFM Heisenberg 
models have with the $Cu-O$ planes in the new ceramic superconducting materials 
\ci{BEMS}, but spin chains and ladders have become rapidly a subject
of autonomous interest in their own.

Back in the early $80$'s Haldane 
\ci{Ha} put forward his by now famous ``conjecture'' according to
which half-odd-integer AFM Heisenberg spin chains should have a gapless
spectrum, and hence be critical at $T=0$ with algebraic decay of
correlations while integer spin chains should be gapped with exponential
decay of correlations. The conjecture is based on the presence, in the
effective action that, in the continuum limit, maps the chain onto a ($1+1$) 
nonlinear Sigma model [$NL\sigma M$] \ci{Ra}, of a topological term 
(basically a Pontrjagin index 
\ci{Ra,Au,Fra,Mo}), that is effective only when the spin is
half-odd-integer (for details see, e.g. 
\ci{AMS}). It was later proved rigorously 
\ci{Za2} that a ($1+1$) $NL\sigma M$ {\it without} topological terms
has a unique ground state and is gapped with an excitation spectrum
consisting of a triplet of massive bosons, as well as that the topological
term makes the model gapless \ci{SR}. 
This implies of course that {\it all} half-odd-integer spin
chains fall into the same universality class of the spin-$1/2$ chain, whose
exact solution is well established by the Bethe-Ansatz 
\ci{Be,Ma}, and that is known to be gapless. Were it not for the
presence of the topological term, the description of half-odd-integer spin
chains in terms of effective nonlinear $\sigma$ model [NL$\sigma$M] actions 
would lead then to
predictions in blatant disagreement with the results of the
Lieb-Schultz-Mattis [LSM] theorem \ci{LSM}, which predicts a (non Goldstone)
gapless sector in the excitation spectrum in the presence of a unique ground
state. All this applies to systems with translationally- (and
parity)-invariant couplings. Gapfull half-odd-integer spin chains can occur
when these invariances are explicitly broken \ci{AMS}. 

The conclusions of LSM do not apply to integer-spin chains, though, and
there a unique ground state {\it and }a gap (the ``Haldane gap'') may well
coexist. Indeed, neutron-scattering measurements on quasi one-dimensional
compounds \ci{RZR} have established the presence of the Haldane gap in a class
of integer-spin systems. Also, a rigorous example of an $S=1$ model with a
gapped unique ground state has been exhibited (among other rigorous results)
in a seminal paper by Affleck et al. \ci{AKLT}.

In more recent years there has arisen a growing interest in the behaviour of
spin chains and/or ladders in an external (uniform and/or staggered 
\ci{Yu}) static magnetic field. Actually, the magnetization curve of the
spin one-half AFM Heisenberg chain had been studied already by Griffiths
back in the early Sixties 
\ci{Gr}. The main result of Ref.\ci{Gr} was that the magnetization $m$ per site, 
as a function of the external field $H$, grows steadily from $H=0$ 
(with a finite slope corresponding to a finite initial susceptibility, 
as expected) to the saturation value of $m=$ $S=1/2$. 
The latter is reached at an upper critical field $H_{c_{2}}$ given by 
(absorbing the Bohr magneton and the spin $g$-factor in the definition of 
the field): $H_{c_{2}}=4SJ=2J$ for $S=1/2$ ($J$ being the AFM exchange constant) 
with infinite slope: $S-m\propto (H_{c_{2}}-H)^{1/2}$. 
A continuous (in the thermodynamic limit) change of the magnetization as a 
function of the applied field implies of course that there are no gaps in the 
spectrum towards excitations from a value of the spin to a neighboring one. 
The $S=1/2$ model remains therefore gapless from $H=0$ up to $H=H_{c_{2}}$, 
where the ground state becomes fully polarized. In this connection, it should be 
mentioned that the {\it opposite} phenomenon, namely a 
{\it field-induced} gap has been observed 
\ci{De} in neutron-scattering experiments on what can be considered
as a (quasi) one-dimensional $S=1/2$ spin system, namely $Cu$ benzoate. It
has however been argued convincingly 
\ci{OA} that this is possible only in the presence of various types
of anisotropies, leading to an effective internal staggered field (see 
\ci{OA} for more details).

As already stated, integer-spin (isotropic) AFM chains are in a gapped
``Haldane'' phase at zero field. This appears to be true 
\ci{PB} also for XXZ models, i.e. for more general model Hamiltonians of the form: 
\begin{equation}  
{\cal H=}J\sum_{i=1}^{N}\left\{\lambda S_{i}^{z}S_{i+1}^{z}+\frac{1}{2}%
(S_{i}^{+}S_{i+1}^{-}+S_{i}^{-}S_{i+1}^{+})\right\} 
\end{equation}  
for a chain of $N$ sites (we don't specify here boundary conditions (see 
\ci{PB} for details)), where $\lambda \geq 0$ is an anisotropy
parameter in a range that includes the (exactly soluble, $\lambda =0$) $XY$
model, the (isotropic, $\lambda =1$) Haldane phase as well as, for $\lambda
>1$, uniaxial Ising-like phases.

Sticking for the time being to the isotropic, $\lambda=1$ case, and
according to the analysis of \ci{Za2}, the lowest excited state 
is a triplet of massive, $S=1$
bosons. In the absence of an external magnetic field the latter are
degenerate and separated from the ground state by a finite energy gap 
$\Delta_{0}.$ The addition to the model of an interaction with an external field
leads to new and interesting phenomena.

As observed by Affleck 
\ci{A91}, the Zeeman splitting due to the field will cause one of the
branches of the spectrum  to lower its energy. The gap will remain
``robust'' (and the magnetization will remain zero) up to a lower critical
field: $H_{c_{1}}=\Delta_{0} $, when the lower branch of the spectrum will cross
the ground state. At this point, a ``Bose condensation of magnons'' 
\ci{OYA} should take place, and the spectrum should become gapless.
This implies of course that the magnetization should remain zero for
fields up to the lower critical field, while it should rise to the
saturation value at some upper critical field  $H_{c_{2}}\geq H_{c_{1}}$. 
The value $H=H_{c_{1}}$ of the field is then the one that leads to the
closure of the Haldane gap in the excitation spectrum. Parenthetically, it
should be remembered that the existence, for integer spins, of a lower
critical field leading to the closure of the Haldane gap had already been
predicted previously by Schulz \ci{Schu1}.

The Lieb-Schultz-Mattis Theorem \ci{LSM}, originally formulated in zero 
external field, has been generalized to both spin chains 
\ci{A88,OYA} and ladders \ci{CHP} in the presence of a field. 
The main results of \ci{OYA} for spin chains (but see also 
\ci{CHP} for a generalization) are that, in the presence of a field:

i) If translational symmetry is {\it not} broken, then the ground state for
a spin-$S$ chain is gapless {\it unless} the magnetization per site $m$
obeys the ``quantization condition'': 
\begin{equation}  
S-m\in Z,\text{ an integer}
\end{equation}  

This implies that gaps in the spectrum, and hence plateaus in the
magnetization curve {\it can} (but in principle need not, of course) be
present when the above ``integer quantization'' condition is met. Moreover:

ii) ``Fractional quantization'' (completing an overall scenario that
resembles closely 
\ci{OYA} that of the (integer and fractional) Quantum Hall Effect),
i.e.: $S-m=$ half-odd-integer, may lead to additional ``fractional''
plateaus accompanied, however, by spontaneous breaking of the translational
symmetry of the ground state.

Although plateaus (both integer and fractional) are not forbidden at
quantized values of the magnetization, whether or not they do actually occur
depends very much on the details of the model. For $S=1$ the only admissible
gaps are at $m=0$ (with a width: $H_{c_{1}}\approx\Delta_{0}$, the Haldane gap)
and (trivially) at $m=S=1$ for $H\geq H_{c_{2}}$, corresponding to the fully
polarized state. In order to have gaps at intermediate values one needs
therefore $S\geq3/2$. The existence of a gap at $m=1/2$ has been proved by
Tasaki \ci{Ta} and numerically found \ci{OYA,ST2} only in the presence of a 
strong single-ion anisotropy. Some numerical \ci{CHP} and experimental 
\ci{Chab} evidence of magnetization plateaus has been reported for 
$S=1/2$ (even-legged) spin ladders.

We will restrict ourselves from now on to the case of $S=1$ AFM isotropic
Heisenberg chains. For these systems the magnetization is supposed to rise
continuosly for $H_{c_{1}}\leq H\leq H_{c_{2}}$ from zero to saturation. The
model being gapless in this range of fields, the transverse spin-spin
correlation function (e.g.: $\langle S_{0}^{x}S_{r}^{x}\rangle $ if the
external field $H$ is in the $z$-direction) is expected to decay
algebraically: $\langle S_{0}^{x}S_{r}^{x}\rangle \propto (-1)^{r}r^{-\eta }$
with some critical exponent $\eta $. 
Affleck \ci{A91} has suggested that the ground state above $H_{c_{1}}$ should
be regarded as a Bose condensate of low-energy bosons and has proposed  
an effective Ginzburg-Landau-type field theory to describe the system near 
the transition. His predictions are that $\eta$ should exhibit the universal
behaviour: $\eta \simeq 1/2$ near $H_{c_{1}}$ and that the induced
magnetization should vanish in a nonanalytic way at $H_{c_{1}}$, namely: 
$m\propto \sqrt{H-H_{c_{1}}\text{ }}$for $H\gtrsim H_{c_{1}}$. 
This prediction seems to be reasonably well supported by numerical findings 
\ci{ST1}. In the same reference it was also inferred from the
numerical data that $\eta \approx 1/2$ \ as well when 
$H\rightarrow H_{c_{2}} $. 
It is unclear whether such a prediction can be actually extracted 
firmly from the available numerical data.

The motivations of the present paper are twofold: we wanted to extend and to
put on firmer bases the existing numerical analyses of the model, which has
proven to be possible by a systematic use of the DMRG that has allowed us
to analyze samples with an higher number of sites (up to $N=80$) than down
hitherto, and to try and understand the Haldane phase in terms of an
appropriate effective field theory.

The plan of the paper is as follows: In Sect. II we give the details of the
numerical procedure followed in the study (at $T=0$) of finite samples and
report the results for the Haldane gap, the critical fields, the
magnetization curve and the critical exponent for the (transverse) spin-spin
correlation function. In Sect. III we discuss, at the mean field level, the
predictions of an effective field theory based on the $O(3)$ nonlinear 
$\sigma$ model (NL$\sigma$M) for the Haldane phase of the chain, comparing
the predictions of the theory with the numerical results. Sect. IV is
devoted to a brief discussion and to the conclusions.

\section{Numerical Results}
 
The Hamiltonian we consider here is:
\begin{equation}  
{\cal H}=J\sum\limits_{\langle ij\rangle }{\bf S}_{i}\cdot {\bf S}_{j}-
H\sum\limits_{i}S_{i}^{z} \label{ham} 
\end{equation}  
where: {\bf S}$_{i}^{2}=S(S+1)$, $S=1$ (we set $\hbar =1$ from now
on), the sums run over a chain of $\ N=L/a$ spins, $L$ being the
length of the chain and $a$ the lattice spacing. $J$ is the AFM coupling
between nearest-neighboring spins and $H$ (which includes factors such as
Bohr magneton, gyromagnetic factor and the like) is a static and uniform
external magnetic field pointing in the $z$-direction. Here and in the rest
of the paper we shall assume periodic boundary conditions (PBC's).

We have employed the standard DMRG algorithm \ci{WH,WN,WW,Spa} 
to calculate the magnetization curve and the static
transverse spin-spin correlation function: 
\begin{equation}  
G^{xx}(r)=\langle S_{r}^{x}S_{0}^{x}\rangle  
\end{equation}  
($r=j$, in units of $a$ with $j$ an integer, $0\leq j\leq N$).
From the latter one can extract both the value of the correlation length
(and hence of the spin gap) in the Haldane phase and, as a function of,
e.g., the magnetization, the critical exponent governing the algebraic
decay of correlations in the gapless phase. We have adopted a finite-chain
``zip'' algorithm which increases considerably the precision of the
resulting data, though of course at the price of an increase in computer
time. In the whole of this Section we measure energies in units of $J$, i.e.
we set $J=1$.

\begin{figure}
\hspace{-0.3cm}
\epsfxsize=4cm \epsffile{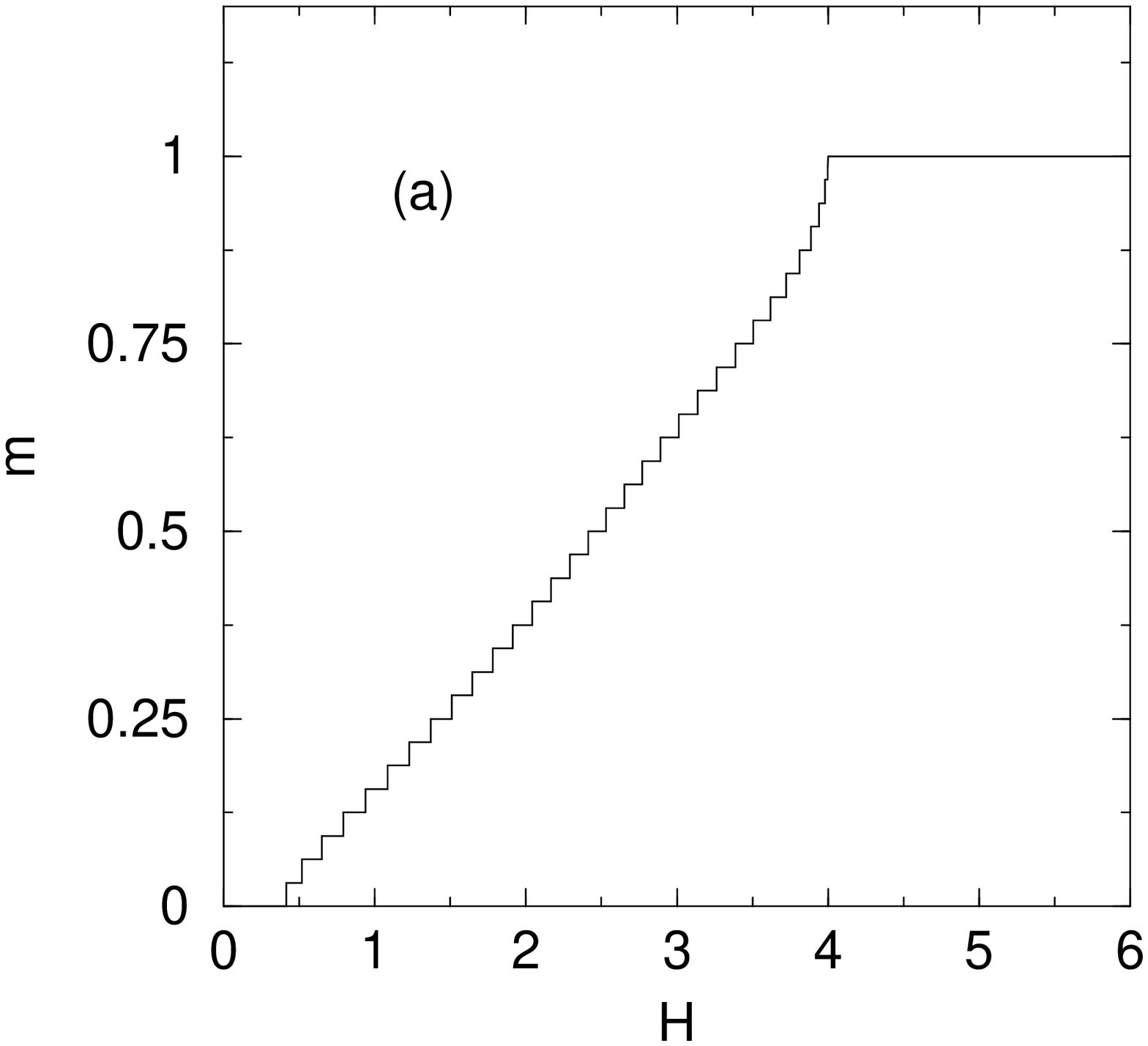}\hfill
\epsfxsize=4cm \epsffile{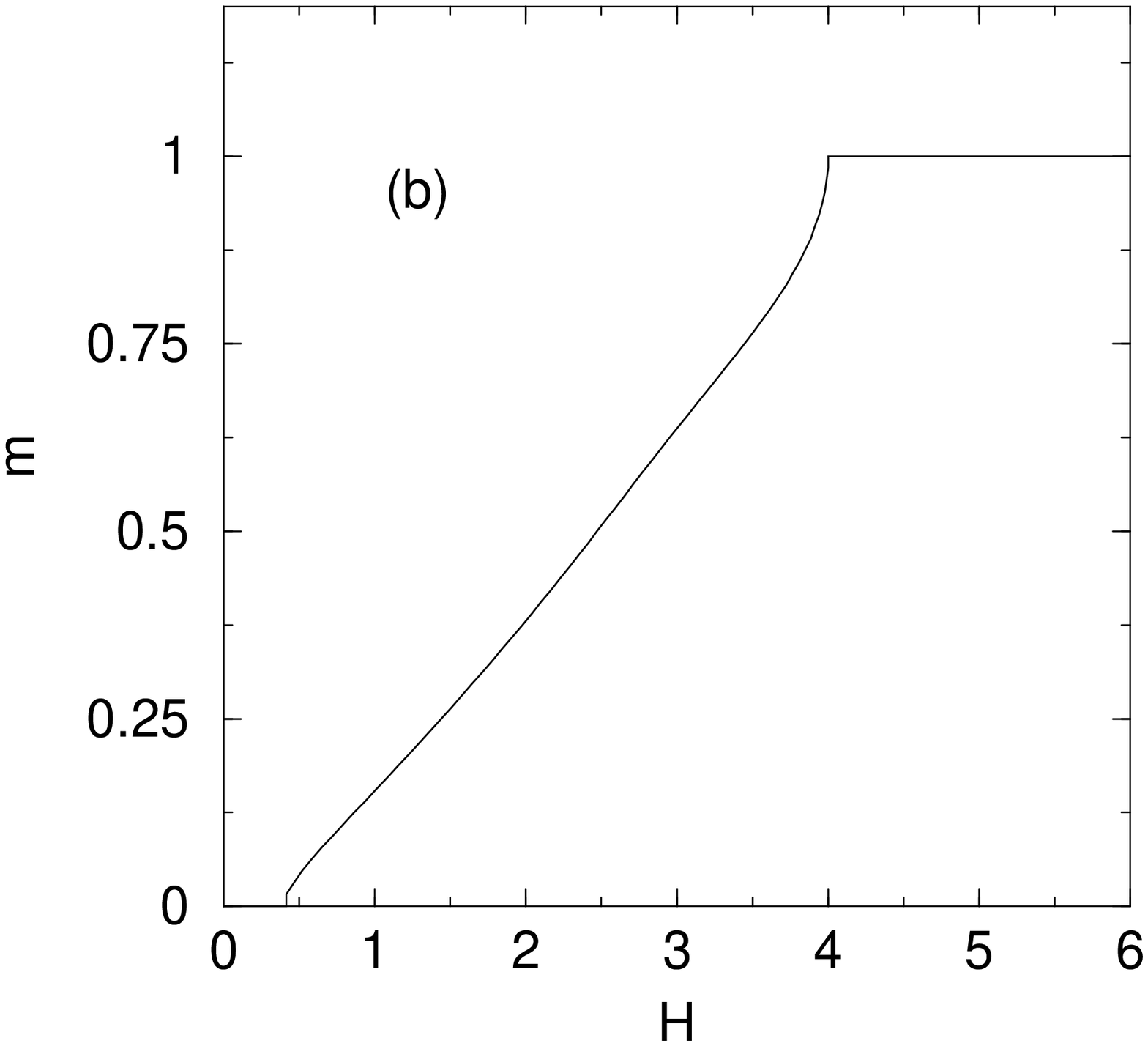}
\narrowtext
\caption[]{a) The hystogram of the magnetization curve for $N=32$ spins. 
b) Continuous interpolation of the data in Fig. 1-a. See text for details.}
\label{fig1}
\end{figure}
The hystogram of the magnetization for $N=32$ spins is reported in 
Fig. 1-a, while Fig. 1-b represents a continuous interpolation of the data. The
curve in Fig. 1-b passes through the midpoints of the plateaus of the 
hystogram and is an indication of what the magnetization curve 
should look like in the thermodynamic limit $N\to \infty .$

The magnetization remains zero up to a lower critical field of 
$H_{c_{1}}=0.412232$ (in our units). 
Identifying $H_{c_{1}}$ with $\Delta_{0} $, $\Delta _{0}$ being the spin gap, 
this is a close upper limit to the $N\to \infty $ extrapolations 
\ci{WH} that give: $\Delta _{0}/J$=$0.40450(2).$ 
Similar results have been reported also by Golinelli et al. \ci{GJL}. 
Above the lower critical field the magnetization increases steadily
until it saturates at the upper critical field $H_{c_{2}}/J=4$ (equal to 
$4S$ for $S=1$). 
From the data of Fig.$1$, it emerges that the numerical analysis reproduces 
in a completely nonambiguous way the singular square-root behaviour 
of the magnetization near the upper critical field. 
The data are also consistent with a similar square-root vanishing 
\ci{A91} of the magnetization when the lower critical field is 
approached from above, but seem to indicate that the low-field singular 
behaviour sets in in a much narrower range of fields than the one near 
$H_{c_{2}}$. 
The values of $(-1)^r G^{xx}(r)$ are plotted in 
Fig. 2 for a chain of $N=64$ spins and $H<H_{c_{1}}$. PBC's make of course the curve 
symmetric around $N/2=32$. 
The data are well fitted, taking PBC's into account, by a curve of the form: 
\begin{equation}  
(-1)^r G^{xx}(r)= A\left\{ \frac{e^{-r/\xi _{0}}}{\sqrt{r}}+\frac{%
e^{-(N-r)/\xi _{0}}}{\sqrt{N-r}}\right\} 
\end{equation}  
 with: 
\begin{equation}  
A=1.07508,\text{ \ }\xi _{0}=6.08304 
\end{equation}  
both independent of the field up to $H=H_{c_{1}}$. 
The form of $G^{xx}(r)$ reported here corresponds to the (symmetrized) 
asymptotic behaviour of the modified Bessel function 
\ci{Grad} $K_{0}(r/\xi ),$ which is the form of the 
spin-spin static correlation function predicted by the continuum theory 
to be discussed in the next Section. Also, defining the spin-wave
velocity $c$ as: $c=\Delta _{0}\xi $ ($\Delta _{0}\xi /\hbar $ if we
reinstate Planck's constant into its proper place) we find 
$c\simeq 2.4$ in units of $Ja$ (or $Ja/\hbar $). 
The continuum theory (see below) predicts instead 
$c=2$ (for $S=1$ and in the same units). The discrepancy may be 
attributed both to the approximations involved in performing the continuum
limit and to finite-size effects that may affect the numerical estimates.

\begin{figure}
\centering
\epsfxsize=7cm \epsffile{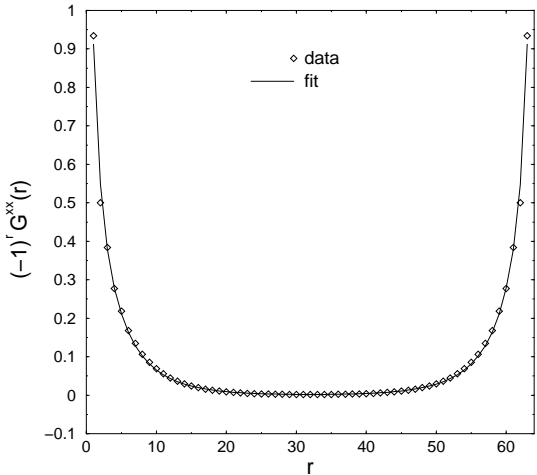}
\narrowtext
\caption[]{A plot of $(-1)^r G^{xx}(r)$ ($r=j$, in units of $a$, 
where $j$ is an integer) for 
a chain of $N=64$ spins and $H<H_{c_1}$. 
The points are the numerical data and the continuous curve is the 
interpolation with the analytic form of Eq.(5).}
\label{fig2}
\end{figure}
For $H_{c_{1}}\leq H\leq H_{c_{2}}$ the correlation function is expected to
exhibit an asymptotic algebraic decay, i.e. to behave asymptotically as: 
\begin{equation}  
(-1)^r G^{xx}(r)\approx \frac{1}{r^{\eta }}+\frac{1}{(N-r)^{\eta }} 
\label{powerlaw}
\end{equation}  

Here too the second term accounts for the PBC's and $\eta =\eta (m).$
According to Moreo\ci{Moreo} and to Sakai and Takahashi\ci{ST1} 
the most efficient way to calculate $\eta$ goes through the 
evaluation of the (staggered) structure factor: 
\begin{equation}  
S_{N}=\sum\limits_{r=1}^{N-1}\exp \{ikr\}G^{xx}(r)\vert_{k=\pi } 
\end{equation}  

A power-law behaviour as in Eq.(\ref{powerlaw}) implies: 
$S_{N}\approx N^{1-\eta }$ for
large $N,$ and one obtains the critical exponent as: 
\begin{equation}  
\eta (m)=1-\frac{\ln \left(S_{N}/S_{M}\right)}
{ \ln \left(N/M\right)} 
\end{equation}  
with the most reliable values being obtained for the highest $M,N$ available.

In Fig.$3$ we report our numerical results for $\eta $ together with some
interpolations. The data have been obtained by employing about eight runs
for each value always using the ``zip '' algorithm twice and 
keeping between $190$ and $234$ states in the reduced density matrix. 
The longest chains studied here have up to $N=80$ spins.

\begin{figure}
\centering
\epsfxsize=7cm \epsffile{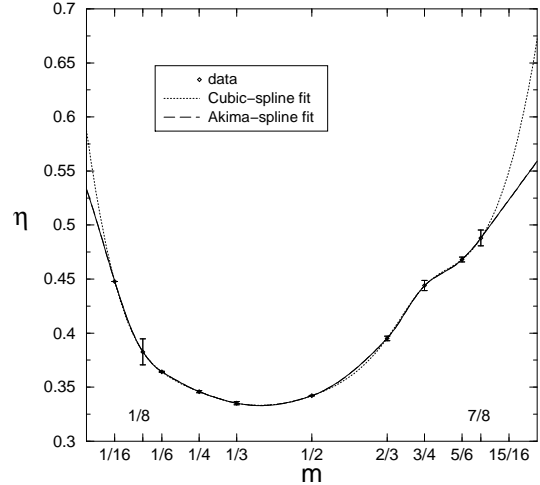}\hfill
\narrowtext
\caption[]{The crtitical exponent $\eta$ (see Eq.(7)) as a function of the 
magnetization for chains up to $N=80$ spins. 
The interpolating curves should be taken mainly as guides for the eye.}
\label{fig3}
\end{figure}
For $m\to 0$ we obtain extrapolated values of $\eta $ slightly above the
theoretical prediction \ci{A91} of $\eta =1/2$. This may be due to
uncertainties in the interpolation scheme and again to finite-size effects.
For $m\to 1$, i.e. near saturation, our data seem to be consistent with
a finite limiting value of $\eta $ at saturation.

Our numerical results indicate that the system remains in the Haldane phase
for fields up to $H_{c_{1}}$, while it becomes (quantum) critical for higher
fields when a nonzero magnetization develops. In the next Section we will
try to set up a consistent field-theoretical description of the Haldane
phase.

\section{Effective Field Theory for the Haldane Phase}

That AFM Heisenberg spin chains can be mapped, in the continuum limit,
into O(3) NL$\sigma$ models was shown for the first time by Haldane 
\ci{Ha} and such ``Haldane maps'' have been widely used since. As the
procedure is by now well known, we summarize here only its main points.

In order to use the partition function as a generating functional for
(Euclidean) Greeen functions, we will replace for the time being the
Hamiltonian of Eq.(\ref{ham}) with:

\begin{equation}  
{\cal H}=J\sum\limits_{\langle ij\rangle }{\bf S}_{i}\cdot {\bf S}_{j}-\sum_{i}{\bf B}_{i}\cdot 
{\bf S}_{i},\text{ }J>0 
\end{equation}  
where ${\bf B}_{i}$ is an external site- and possibly time-dependent
magnetic field. Here too PBC's will be assumed throughout.

The (canonical) partition function can be written in a standard manner as a
path integral over spin coherent states as: 
\begin{equation}  
{\cal Z}=\int [{\cal D}{\bf \Omega }]\delta ({\bf \Omega }^{2}-1)\exp \{-%
{\cal S}_{E}\} 
\end{equation}  
where the (Euclidean) action is given by: 
\begin{equation}  
{\cal S}_{E}={\cal S}_{WZ}+\int\limits_{0}^{\beta }{\cal H}[{\bf \Omega }(\tau
)]d\tau 
\end{equation}   
${\bf \Omega }$ is a classical vector field of unit magnitude,
\begin{equation}  
{\cal H}[{\bf \Omega }(\tau )]=JS^{2}\sum\limits_{\langle ij\rangle }{\bf \Omega }_{i}(\tau
)\cdot {\bf \Omega }_{j}(\tau )-S\sum_{i}{\bf B}_{i}(\tau )\cdot {\bf \Omega 
}_{i}(\tau ) 
\label{hamom}
\end{equation}  
and ${\cal S}_{WZ}$ is the Wess-Zumino part of the action given by: 
\begin{equation}  
S_{WZ}= i s \sum_i \int_{0}^{\beta}\,  
d\tau [1-\cos\theta_i(\tau)]\,{\dot\phi}_i(\tau)
\end{equation}  
where $\theta_i$ and $\phi_i$ are the polar coordinates of the parametrization 
of ${\bf \Omega }_{i}$.
Following Haldane, we decompose ${\bf \Omega }_{i}$ into a staggered 
N\'eel field ${\bf n}_{i}$ and a fluctuation field ${\bf l}_{i}$ as: 
\begin{equation}  
{\bf \Omega }_{i}=(-1)^{i}{\bf n}_{i}\sqrt{1-a^{2}{\bf l}_{i}^{2}}+a{\bf l}%
_{i} \label{ansatz}
\end{equation}  
and the constraint ${\bf \Omega }_{i}^{2}=1$ is enforced by: 
${\bf n}_{i}^{2}=1$ and ${\bf n}_{i}\cdot {\bf l}_{i}=0.$

Performing then the continuum limit followed by a gradient expansion and 
integrating out the fluctuation field, one ends up with the following 
expression for the partition function expressed as a path integral over the 
N\'eel field alone:
\begin{equation}  
{\cal Z}=\int [{\cal D}{\bf n}]\delta \lbrack {\bf n}^{2}-1]\exp [-{\cal S}%
_{E}] 
\end{equation}  
with the Euclidean action: 
\begin{equation}  
{\cal S}_{E}=\int\limits_{0}^{L}dx\int\limits_{0}^{\beta }d\tau {\cal L}%
_{E}(x,\tau ) 
\end{equation}  
and
\begin{eqnarray}
{\cal L}_{E}(x,\tau )&=&\frac{1}{2g}\left\{\frac{1}{c}|\partial _{\tau }
{\bf n}|^{2}+c|\partial _{x}{\bf n}|^{2}\right\}-\frac{1}{2gc}\left({\bf B}
^{2}-({\bf B\cdot n})^{2}\right) \nonumber \\
& & + \; \frac{i}{gc}{\bf B\cdot n\times }\partial _{\tau }{\bf n} 
\end{eqnarray}
where $g=2/S$ and $c=2JSa$ are the bare coupling and spin-wave velocity, 
respectively. 
This form of the Euclidean Lagrangian has already been derived by
other authors \ci{Mi,LN,AL}. 
The last term in ${\cal L}_{E}(x,\tau )$ is basically geometric 
in nature, being linear in time derivatives, and comes from the 
variation of the Wess-Zumino part of the action. 
The Berry phase instead, that comes also from the Wess-Zumino
action, can be safely neglected as we are dealing here with integer spins.

An additional term that is present, e.g., in the derivation of the effective
action in Ref.\ci{AL} gives rise, in the continuum limit, to an 
integrated total (space) derivative that, as a consequence of the assumed 
PBC's, does not contribute to the overall Euclidean action. 
However, it can affect in a
crucial way the evaluation of the spin Green functions. Indeed, using 
Eq.(\ref{ansatz}), the spin Green functions are given in the exact 
(lattice) version of the theory by \ci{SA}:
\begin{eqnarray}
G_{ij}^{\alpha\beta}(\tau) &=&\left\langle 
T_{\tau}\{S_{i}^{\alpha}(\tau)S_{j}^{\beta}(0)\}\right\rangle\nonumber \\
&=&S^{2}(-1)^{i-j}\left\langle T_{\tau}\{n_{i}^{\alpha}(\tau)n_{j}^{\beta}(0)\}
\right\rangle\nonumber \\ 
&+& S^{2}a^{2}\left\langle 
T_{\tau}\{l_{i}^{\alpha}(\tau)l_{j}^{\beta}(0)\}\right\rangle
\label{correl}
\end{eqnarray}
to leading order in $\mathbf{l}$ and with the expectation values of the
cross terms vanishing by symmetry. According to the analysis of the first
of Refs.\ci{SA} the second term in Eq.(\ref{correl}) 
(a two-magnon contribution)
costitutes a minor correction to the dominant, staggered part and, in the
static (i.e. equal-time) limit it decays asymptotically faster than the
staggered part. So, we will approximate the spin Green functions as:
\begin{equation}
G_{ij}^{\alpha\beta}(\tau)\simeq
S^{2}(-1)^{i-j}\left\langle T_{\tau}\{n_{i}^{\alpha}(\tau)n_{j}^{\beta}(0)\}
\right\rangle
\end{equation}
having in mind that, due to the staggering prefactor, the small-$q$
behaviour in Fourier space of the Green functions for the $\mathbf{n}$-field
will correspond to that of the spin Green functions near $q=\pi$.

One may wish also, for reasons of internal consistency, to derive the Green
functions (actually the truncated ones) directly from the partition function
via functional differentiation, i.e. as:
\begin{equation}
G_{ij}^{\alpha\beta}(\tau)=\frac{\delta^{2}\mathcal{Z}}{\delta B_{i}^{\alpha
}(\tau)\delta B_{j}^{\beta}(0)}
\label{corr2}
\end{equation}
This can be done safely, however, only {\it before} the continuum limit is 
taken, as can be seen easily from a careful analysis of the Zeeman term
in the Hamiltonian (\ref{hamom}). 
With the substitution (\ref{ansatz}) the latter becomes:

\begin{eqnarray}
S\sum\limits_{i}\mathbf{B}_{i}(\tau)\cdot\mathbf{\Omega}_{i}(\tau
)&=&S\sum\limits_{i}\left\{(-1)^{i}\mathbf{B}_{i}(\tau)\cdot\mathbf{n}
_{i}(\tau)\right. \nonumber \\
&+&\left. a \mathbf{B}_{i}(\tau)\cdot\mathbf{l}_{i}(\tau) \right\}
\label{zeeman}
\end{eqnarray}
and it is pretty obvious that the double functional differentiation of 
Eq.(\ref{corr2}) will reproduce the r.h.s. of Eq.(\ref{correl}). 
The first term in Eq.(\ref{zeeman}) is
however precisely the one that maps, in the continuum limit, into the
already-mentioned irrelevant (in that limit) total derivative term. So, if
functional differentiations were performed (erroneously) {\it after} taking the
continuum limit they would only reproduce the nonstaggered part (with the
site indices $i,j$ replaced by continuos variables, of course) of the Green
functions (\ref{correl}). 
This we have checked also by direct calculation, performing
the functional differentiation both before and after having done the
Gaussian integration over the fluctuation field.

Returning now to the evaluation (in the continuum limit now) of the 
partition function, the NL$\sigma$M constraint can be implemented by 
re-expressing the partition function as: 
\begin{equation}  
{\cal Z}=\int[{\cal D}{\bf n}]\left[\frac{{\cal D}\lambda}{2\pi}\right]
\exp[-{\cal S}_{eff}] 
\end{equation}  
where now: 
\begin{equation}  
{\cal S}_{eff}=
\int\limits_{0}^{L}dx\int\limits_{0}^{\beta}d\tau {\cal L}_{eff}(x,\tau) 
\end{equation}  
and 
\begin{equation}  
{\cal L}_{eff}(x,\tau)={\cal L}_{E}(x,\tau)-i\lambda(x,\tau )({\bf n}%
^{2}(x,\tau)-1) 
\end{equation}  
with $\lambda (x,\tau )$ an auxiliary (real) Lagrange multiplier field. The
total effective action is now quadratic in the field ${\bf n}$ and is given
by: 
\begin{eqnarray}  
S_{eff}&=&\int\limits_{0}^{L}dx\int\limits_{0}^{\beta }d\tau
\int\limits_{0}^{L}dx^{\prime }\int\limits_{0}^{\beta }d\tau ^{\prime}
\nonumber\\
& &\left\{n^{\alpha }(x,\tau )(K^{-1})^{\alpha \gamma}(x\tau ;
x^{\prime }\tau^{\prime })n^{\gamma }(x^{\prime },\tau ^{\prime})\right\}  
\nonumber\\
&+& \int\limits_{0}^{L}dx\int\limits_{0}^{\beta }d\tau 
\left\{-\frac{1}{2gc}{\bf B}^{2}(x,\tau )+i\lambda (x,\tau )\right\} 
\end{eqnarray}  
where:
\begin{eqnarray}
&&(K^{-1})^{\alpha \gamma }(x\tau ;x^{\prime }\tau ^{\prime }) =
\nonumber\\
&&\left\{-\frac{1}{2g}[\frac{1}{c}
\partial _{\tau }^{2}+c\partial_{x}^{2}]\delta^{\alpha \gamma }
+\frac{1}{2gc}B^{\alpha }(x,\tau )B^{\gamma }(x^{\prime}\tau ^{\prime })
\right.\nonumber\\
& & \left. +\frac{i}{gc}\epsilon ^{\alpha \gamma 
\delta }B^{\delta }(x,\tau )\partial_{\tau }-i\lambda (x,\tau )
\delta ^{\alpha \gamma }\right\}\delta 
(x-x^{\prime })\delta (\tau -\tau ^{\prime })
\end{eqnarray}
and we can proceed to integrating out ${\bf n}$, leaving a resulting
effective action depending only on ${\bf B}$ and the auxiliary field 
$\lambda.$ 
The integration involves the evaluation of the functional
determinant of $K$, and this is most easily done in Fourier space.

Fourier transforms \ will be defined according to: 
\begin{eqnarray}
B^{\alpha }(x,\tau )&=&\frac{1}{\beta L}\sum\limits_{q,n}\exp [i(qx-\Omega
_{n}\tau )]\widetilde{B}^{\alpha }(q,n) \\
\widetilde{B}^{\alpha }(q,n)&=&\int\limits_{0}^{L}dx
\int\limits_{0}^{\beta}d\tau \exp 
[-i(qx-\Omega _{n}\tau )]B^{\alpha }(x,\tau )
\end{eqnarray}
and similarly for the components of ${\bf n}$ and for the field $\lambda.$
The frequencies here are Matsubara-Bose frequencies: 
$\Omega_{n}=2n\pi/\beta. $
Note that reality of ${\bf B}$, ${\bf n}$ and $\lambda$ implies: 
\begin{equation}  
\widetilde{B}^{\alpha}(-q,-n)=\{\widetilde{B}^{\alpha}(q,n)\}^{\ast} 
\end{equation}  
as well as: 
\begin{equation}  
\widetilde{n}^{\alpha}(-q,-n)=\{\widetilde{n}^{\alpha}(q,n)\}^{\ast} 
\end{equation}  
and
\begin{equation}  
\widetilde{\lambda}(-q,-n)=\{\widetilde{\lambda}(q,n)\}^{\ast} 
\end{equation}  

Some long algebra leads then to:
\begin{eqnarray}  
{\cal S}_{eff}&=&\sum\limits_{q,q^{\prime }}\sum\limits_{n,n^{\prime }}
\widetilde{n}^{\alpha }(q,n)^{\ast }
(K^{-1})^{\alpha \gamma }(q,n;q^{\prime}n^{\prime })
\widetilde{n}^{\gamma }(q^{\prime },n^{\prime })
\nonumber\\
&&-\frac{1}{2gc}\frac{1}{\beta L}\sum_{q,n,\alpha }
\big\vert\widetilde{B}^{\alpha }(q,n)\big\vert^{2}+i%
\widetilde{\lambda }(0,0) 
\end{eqnarray}  
where 
\begin{eqnarray}
&&(K^{-1})^{\alpha \gamma }(q,n;q^{\prime },n^{\prime }) =\delta
_{nn'}\delta _{qq^{\prime }}\delta ^{\alpha \gamma }\frac{1}{2gc}[\Omega
_{n}^{2}+c^{2}q^{2}] \nonumber\\
&&+\frac{1}{2gc}\frac{1}{(\beta L)^{3}}\sum\limits_{km}\widetilde{B}^{\alpha
}(k+q,m+n)\widetilde{B}^{\gamma }(k+q^{\prime },m+n^{\prime }) \nonumber\\
&&+\frac{1}{gc}\frac{1}{(\beta L)^{2}}\Omega_{n^{\prime }}
\epsilon^{\alpha\gamma \delta}\widetilde{B}^{\delta }
(q-q^{\prime },n-n^{\prime })\nonumber\\
&&-i\delta^{\alpha \gamma }\frac{1}{(\beta L)^{2}}
\widetilde{\lambda }(q-q^{\prime},n-n^{\prime })
\end{eqnarray}

It is clear that the Fourier transforms of the (nonlocal in the presence
of a general external field and/or Lagrange multiplier $\lambda )$ Green
functions of the {\bf n} field are given by: 
\begin{equation}  
\langle n^{\alpha }(q,n)n^{\gamma }(q^{\prime },n^{\prime })\rangle =K^{\alpha \gamma
}(q,n;q^{\prime },n^{\prime }) 
\end{equation}

We can perform now the Gaussian integration over the field ${\bf n}$ with
the result: 
\begin{equation}  
{\cal Z=}\int [{\cal D}\lambda ]\exp \{-S[\lambda ;{\bf B}]\} 
\end{equation}  
where 
\begin{equation}  
S[\lambda ;{\bf B}]={\rm Tr}\ln (K^{-1})-\frac{1}{2gc}\frac{1}{\beta L}%
\sum_{q,n,\alpha }|\widetilde{B}^{\alpha }(q,n)|^{2}+i\widetilde{\lambda }%
(0,0); 
\end{equation}  
\begin{equation}  
{\rm Tr}\ln (K^{-1})\equiv\sum_{q,n,\alpha }
\{\ln (K^{-1})\}^{\alpha \alpha }(q,n;q,n) 
\end{equation}  

From now on we will limit ourselves to the case of a static and uniform
magnetic field, i.e. we will set:
\begin{equation}  
\widetilde{B}^{\alpha }(q,n)=\beta L\delta _{q,0}\delta _{n,0}H^{\alpha } 
\end{equation}  
In that case: 
\begin{eqnarray}  
(K^{-1})^{\alpha \gamma }(q,n;q^{\prime },n^{\prime })&=& 
(K_{d}^{-1})^{\alpha\gamma }(q,n;q^{\prime },n^{\prime })\nonumber\\
&-&i\delta ^{\alpha \gamma }\frac{1}{(\beta L)^{2}}
\widetilde{\lambda }(q-q^{\prime },n-n^{\prime }) 
\end{eqnarray}  
where $K_{d}$ is the diagonal (in Fourier space) kernel given by: 
\begin{equation}  
K_{d}^{-1}(q,n;q^{\prime },n^{\prime })=\frac{1}{\beta L}\delta _{nn`}\delta
_{qq`}{\cal G}_{0}^{-1}(q,n) 
\end{equation}  
with: 
\begin{eqnarray}  
({\cal G}_{0}^{-1})^{\alpha \gamma }(q,n)&=&
\delta ^{\alpha \gamma }\frac{1}{2gc}[\Omega_{n}^{2}+c^{2}q^{2}]
\nonumber\\
&+&\frac{1}{2gc}H^{\alpha }H^{\gamma }
+\frac{1}{gc}\Omega_{n}\epsilon^{\alpha \gamma \delta }H^{\delta } 
\end{eqnarray}  

Looking for the saddle point(s) of the effective action, we find that there
exist static and uniform ones of the form: 
\begin{equation}  
\widetilde{\lambda }(q,n)=\lambda \beta L\delta _{q,0}\delta _{n,0} 
\end{equation}  
implying: 
\begin{equation}  
(K^{-1})^{\alpha \gamma }(q,n;q^{\prime },n^{\prime })=\frac{1}{\beta L}%
\delta _{nn`}\delta _{qq`}{\cal G}^{-1}(q,n) 
\end{equation}  
and: 
\begin{equation}  
({\cal G}^{-1})^{\alpha \gamma }(q,n)=({\cal G}_{0}^{-1})^{\alpha \gamma
}(q,n)-i\lambda \delta ^{\alpha \gamma } 
\end{equation}  
At the saddle point the Green functions are diagonal (in Fourier space) and
given by: 
\begin{equation}  
\langle n^{\alpha }(q,n)n^{\gamma }(-q,-n)\rangle ={\cal G}^{\alpha \gamma }(q,n) 
\end{equation}  
The saddle-point equation reads then: 
\begin{equation}  
1=\frac{1}{\beta L}\sum_{q,n}{\rm Sp}\{{\cal G}(q,n)\} 
\label{saddle}
\end{equation}  
where ``Sp'' stands for a trace over the vector indices only.
Frequency summations of the form: 
\begin{equation}  
\frac{1}{\beta }\sum\limits_{n}F(\Omega _{n}) 
\end{equation}   
are performed according to a standard recipe as:
\begin{equation}  
\frac{1}{\beta }\sum\limits_{n}F(\Omega _{n})=\oint\limits_{\Gamma }F(z)\rho
(z)\frac{dz}{2\pi } 
\end{equation}  
where the contour $\Gamma $ goes from $-\infty -i0^{+}$ to $+\infty -i0^{+}$
and backwards from $+\infty +i0^{+}$ to $-\infty +i0^{+}$and:
\begin{equation}  
\rho (z)=\left\{\exp [i\beta z]-1\right\}^{-1} 
\end{equation}  

is a meromorphic function having simple poles at: $z=\Omega _{n}=2\pi
n/\beta $ with residue $(i\beta )^{-1}$. Closing the first integral in the
lower half-plane and the second \ one in the upper half-plane the resulting
integrals can be evaluated with the aid of the residue theorem.

Let's begin by examining the zero-field case: $H^{\alpha }=0.$
Then: 
\begin{equation}  
({\cal G}^{-1})^{\alpha \gamma }(q,n)=\delta ^{\alpha \gamma }\left\{
\frac{1}{2gc}\{\Omega _{n}^{2}+c^2 q^{2}\}-i\lambda \right\} 
\end{equation}  
and the saddle-point equation becomes: 
\begin{equation}  
1=\frac{1}{L}\sum\limits_{q}\frac{1}{\beta}\sum\limits_{n}
\frac{6gc}{\Omega_{n}^{2}+c^{2}(q^{2}+\xi^{-2})};
\text{ }\xi^{-2}\equiv \frac{-2i\lambda g}{c} 
\end{equation}  

Under the assumption that $\xi^{-2}$ be  positive (i.e. that the saddle
point be actually purely imaginary and in the upper half-plane), 
the integrand has simple poles at $z=\pm ic\sqrt{q^{2}+\xi^{-2}}$. Hence: 
\begin{eqnarray}  
\frac{1}{\beta}&&\sum\limits_{n}
\frac{6gc}{\Omega_{n}^{2}+c^{2}(q^{2}+\xi^{-2})}= \nonumber\\
&&\qquad\frac{3g}{\sqrt{q^{2}+\xi^{-2}}}\coth\left\{\frac{1}{2}\beta c\sqrt{
q^{2}+\xi^{-2}}\right\} 
\end{eqnarray}  

In the (thermodynamic) limit $L\to \infty $ the momentum integral
requires the introduction of \ an ultraviolet cutoff in momentum space and
we obtain: 
\begin{equation}  
1=\frac{3g}{2\pi }\int\limits_{0}^{\Lambda }\frac{dq}{\sqrt{q^{2}+\xi ^{-2}}}
\coth \left\{\frac{1}{2}\beta c\sqrt{q^{2}+\xi ^{-2}}\right\} 
\end{equation}  
and, in the zero-temperature limit $\beta \to \infty $: 
\begin{equation}  
1=\frac{3g}{2\pi }\int\limits_{0}^{\Lambda }\frac{dq}{\sqrt{q^{2}+\xi ^{-2}}}
=\frac{3g}{2\pi }\ln \left\{\Lambda \xi +\sqrt{1+(\Lambda \xi )^{2}}\right\} 
\label{limit}
\end{equation}  
($3g/2\pi =3/\pi S$ in our case) leading to:
\begin{equation}  
\xi _{0}\equiv \xi (H=0)=\frac{1}{\Lambda }\sinh (2\pi /3g) 
\end{equation}  
and to the diagonal propagator: 
\begin{eqnarray}  
{\cal G}^{\alpha \gamma }(q,n)&=&\delta ^{\alpha \gamma }{\cal G}(q,n); \\
{\cal G}(q,n)&=&\frac{2gc}{\Omega _{n}^{2}+c^{2}(q^{2}+\xi _{0}^{-2})} 
\end{eqnarray}  
The ultraviolet cutoff will be taken here as a free parameter to be
determined by fitting the results to the existing data for the spin gap.
At the saddle point the poles of the Green functions (that are all equal and
diagonal) are a degenerate triplet of massive excitations with energy: 
\begin{equation}  
\epsilon =\epsilon _{0}(q)=c\sqrt{q^{2}+\xi _{0}^{-2}} 
\end{equation}  
and a zero-field gap: 
\begin{equation}  
\Delta_0 =c\xi _{0}^{-1}\approx \Lambda \exp \{-\pi S/3\} 
\end{equation}  
in the large $S$ limit. This is in agreement with the known exact results
on the model \ci{Za2} and with established estimates of the 
behaviour of the gap for large $S$ \ci{AMS}.
Explicitly, performing the frequency summation, the static propagator is
given by: 
\begin{equation}  
{\cal G}(q)=\frac{1}{\beta }\sum\limits_{n}{\cal G}(q,n)=\frac{g}{\sqrt{%
q^{2}+\xi _{0}^{-2}}} 
\end{equation}  
and, in real space: 
\begin{eqnarray}  
{\cal G}(x)&=&\int \frac{dq}{2\pi }\exp \{iqx\}{\cal G}(q)= \nonumber\\
&&\sqrt{\frac{2}{\pi }}K_{0}(|x|/\xi _{0})
\approx \exp \{-|x|/\xi _{0}\}/\sqrt{|x|} 
\end{eqnarray}  
for $|x|\gg \xi _{0}$, in agreement with the results of Sect. II and with
previous predictions \ci{A91}. Therefore $\xi _{0}$ can be
identified with the zero-field correlation length.

In the general case, fixing the direction of the field along the $z$-axis: 
$H^{\alpha }=H\delta ^{\alpha 3}$, the matrix of the inverse propagators has
the form: 
\begin{equation}  
{\cal G}^{-1}=\left( 
\begin{array}{ccc}
A & C & 0 \\ 
-C & A & 0 \\ 
0 & 0 & B 
\end{array}\right)
\end{equation}  
where: 
\begin{eqnarray}
A& =&A(q,\Omega _{n})=\frac{1}{2gc}\{\Omega _{n}^{2}+c^{2}q^{2}\}-i\lambda \\
B& =&B(q,\Omega _{n})=A+\frac{H^{2}}{2gc} \\
C& =&C(q,\Omega _{n})=\frac{1}{gc}H\Omega _{n}
\end{eqnarray}
Hence: 
\begin{equation}  
{\cal G}= \left( 
\begin{array}{ccc}
\frac{A}{A^{2}+C^{2}}\phantom{\Bigl(} & -\frac{C}{A^{2}+C^{2}} & 0 \\ 
\frac{C}{A^{2}+C^{2}}\phantom{\Bigl(} & \frac{A}{A^{2}+C^{2}} & 0 \\ 
0 & 0 & \frac{1}{B}\phantom{\Bigl(}
\end{array}\right)
\end{equation}  
and: 
\begin{equation}  
{\rm Sp}({\cal G})=\frac{2A}{A^{2}+C^{2}}+\frac{1}{B} 
\label{sp1}
\end{equation}  
 
Proceeding as before in the evaluation of the frequency sums, we find: 
\begin{eqnarray}  
&&\frac{1}{\beta }\sum\limits_{n}B^{-1}(q,\Omega _{n})=\nonumber\\
&&\frac{g}{\sqrt{q^{2}+\xi ^{-2}+(H/c)^{2}}}\coth 
\left\{\frac{\beta c}{2}\sqrt{q^{2}+\xi
^{-2}+(H/c)^{2}}\right\}
\end{eqnarray}  
and: 
\begin{eqnarray}
&&\frac{1}{\beta }\sum\limits_{n}\frac{2A}{A^{2}+C^{2}}=\nonumber\\
&&\frac{g}{\sqrt{q^{2}+\xi ^{-2}+(H/c)^{2}}}\left\{\coth 
\left[\frac{\beta }{2}(c\sqrt{q^{2}+\xi ^{-2}+(H/c)^{2}}+H)\right] 
\right. \nonumber\\
&&\left. +\coth \left[\frac{\beta }{2}(c\sqrt{q^{2}+
\xi^{-2}+(H/c)^{2}}-H)\right]\right\}
\label{sp2}
\end{eqnarray}
For $H\to 0$ we recover the previous case. In the zero-temperature limit
the self-consistency equation reduces to: 
\begin{equation}  
1=3g\frac{1}{L}\sum\limits_{q}\frac{1}{\sqrt{q^{2}+\xi ^{-2}+(H/c)^{2}}}
\end{equation}  
Letting $L\to \infty $ and introducing the same ultraviolet cutoff $%
\Lambda $ as before we find: 
\begin{equation}  
1=\frac{3g}{2\pi }\int\limits_{0}^{\Lambda }\frac{dq}{\sqrt{q^{2}+\xi
^{-2}+(H/c)^{2}}}
\end{equation}  
Comparison with Eq.(\ref{limit}) tells us immediately that the value of $%
\xi ^{-2}$ \ for $H\neq 0$ is connected with that at $H=0$ by: 
\begin{equation}  
\xi ^{-2}(H)+(H/c)^{2}=\xi_0^{-2}
\label{xi}
\end{equation}  
which means that $\xi ^{-1}$ will vanish at a lower critical field $H_{c_{1}}
$ given by:
\begin{equation}  
H_{c_{1}}=c\,\xi_0^{-1}=\Delta _{0}
\end{equation}  
Therefore $H_{c_{1}}$ will be equal to the zero-field gap, and the
divergence of $\xi $ for $H\to H_{c_{1}}$ will \ be the signal of  a
phase transition at $H=H_{c_{1}}$.Explicitly: 
\begin{equation}  
\xi (H)=\frac{c}{\sqrt{H_{c_{1}}^{2}-H^{2}}}
\label{xic}
\end{equation}  

Notice, however, that at $T>0$ $\xi $ can never diverge, since for $\xi ^{-2}=0$
one of the hyperbolic cotangents would develop an infrared singularity which
is not present at $T=0$ (strictly).

We  turn now to the Green functions. When evaluated at the saddle point,
besides the diagonal ones, there appear also Green functions correlating the 
$x$ and $y$ components of the $NL\sigma M$ field ${\bf n}$. We have: 
\begin{eqnarray*}
 \langle n^{x}(q,n)n^{x}(-q,-n)\rangle &=&
 \langle n^{y}(q,n)n^{y}(-q,-n)\rangle =\frac{A}{A^{2}+C^{2}} \\
 \langle n^{x}(q,n)n^{y}(-q,-n)\rangle 
 &=&-\langle n^{y}(q,n)n^{x}(-q,-n)\rangle =-\frac{C}{A^{2}+C^{2}}
\end{eqnarray*}
while: 
\begin{equation}  
\langle n^{z}(q,n)n^{z}(-q,-n)\rangle =\frac{1}{B}
\end{equation}  

Interchange of the $x$ and $y$ directions in the internal space can be
achieved by a rotation of $\pi $ around, say, the $x=y$ axis.  
This will result in a symmetry only if one sends simultaneously 
$h\rightarrow -h$, and
this explains the way the off-diagonal Green functions coupling $n^{x}$ and 
$n^{y}$ are related and can be deduced from one another.

Notice that, being odd in frequency, the static (i.e., equal-time)
off-diagonal correlation functions: 
$\langle n^{x}(q,\tau )n^{y}(-q,\tau )\rangle =
1/ \beta \sum\nolimits_{n}\langle n^{x}(q,n)n^{y}(-q,-n)\rangle 
=-\langle n^{y}(q,\tau)n^{x}(-q,\tau )\rangle $ 
will vanish identically, as they should. There are
therefore dynamical but not static off-diagonal correlations.

Turning now to the elementary excitation spectrum, we observe that the
degenerate triplet of excitations at $H=0$ gets splitted into a 
longitudinal branch with energy 
$\epsilon =\epsilon _{0}(q)$ and two transverse branches with energies:
\begin{equation}  
\epsilon =\epsilon _{\pm }(q)=\sqrt{c^{2}\xi _{0}^{-2}+c^{2}q^{2}}\pm H
\label{eps}
\end{equation}  
and gaps $\Delta _{0}=c\xi _{0}^{-1}$ and 
$\Delta _{\pm }=c\xi _{0}^{-1}\pm H$, which seems to be appropriate \ci{A91} 
for a triplet of $S=1$ magnons in an external field. 
The gap of the lower branch will vanish linearly as $H$ approaches $H_{c_{1}}$.
Thus one mode will go ``soft'' at $H_{c_{1}}$, and this is again a signal 
of instability. This result seems to be in complete
agreement with the picture put forward some time ago by Affleck \ci{A91} 
according to which the transition at $H_{c_{1}}$ is due to a 1D 
Bose condensation of (soft) magnons. 
As it stands, however, the present
mean-field theory cannot be extended in a straightforward way beyond the
lower critical field $H_{c_{1}}$.
The static correlation functions (to be compared with those calculated 
numerically in Sect. II) can be derived immediately from the Green 
functions.
In particular, for the $n^{x}-n^{x}$ correlation function we obtain: 
\begin{equation}  
\langle n^{x}(q,\tau )n^{x}(-q,\tau )\rangle =\frac{g}{\sqrt{q^{2}+\xi ^{-2}+(h/c)^{2}}}%
\equiv \frac{g}{\sqrt{q^{2}+\xi _{0}^{-2}}}
\end{equation}  
i.e. the static correlation function will be the same as in the zero-field
limit, with the same correlation length. Therefore, at $H\neq 0$ the
physical gaps and the ``gaps'' in the static correlation function will have
to be treated as related but distinct objects, coinciding only in the
zero-field limit.

We turn now to the calculation of the magnetization.
The Gibbs free energy $G$ is given by: $G=\beta ^{-1}S$ where, at the 
saddle point: 
\begin{equation}  
S=Tr\ln \{K^{-1}\}-\frac{1}{2gc}\beta L\Delta ^{2} 
\end{equation}  
and: 
\begin{equation}  
\Delta ^{2}=c^{2}\xi ^{-2}+H^{2}
\label{delta} 
\end{equation}  
Notice that (cfr.Eq.(\ref{xi})) at $T=0$: 
$\Delta ^{2}\equiv \Delta _{0}^{2}$ and
is field-independent, but it need not be so at finite temperatures.
Explicitly, and again at the saddle point: 
\begin{equation}
(K^{-1})_{s.p.}^{\alpha \gamma }(qn,q^{\prime }n^{\prime })=
\frac{1}{\beta L}\delta _{n,n^{\prime }}
\delta _{q,q^{\prime }}({\cal G}^{-1})^{\alpha\gamma }(q,n) 
\end{equation}
\begin{eqnarray}
({\cal G}^{-1})^{\alpha \gamma }(q,n)&=&\delta ^{\alpha \gamma }
\frac{1}{2gc}\left\{\Omega _{n}^{2}+c^{2}(q^{2}+\xi ^{-2})\right\}\nonumber\\
&&\qquad\frac{1}{2gc}\delta ^{\alpha \gamma }
\delta ^{\alpha ,3}H^{2}+\frac{H}{gc}\Omega_{n}\epsilon^{\alpha \gamma 3}
\nonumber\\
&=&\delta ^{\alpha \gamma }\frac{1}{2gc}
\left\{\Omega_{n}^{2}+c^{2}q^{2}+\Delta ^{2}-H^{2}(1-\delta ^{\alpha 3})
\right\}\nonumber\\
&&\qquad +\frac{H}{gc}\Omega _{n}\epsilon^{\alpha \gamma 3}
\end{eqnarray}
It follows then that the total magnetization is given by: 
\begin{equation}  
M=-\left(\frac{\partial G}{\partial H}\right)_{\beta }=-\frac{1}{\beta }
\sum\limits_{q,n}
{\rm Sp}\left\{{\cal G}\frac{\partial {\cal G}^{-1}}{\partial H}
\right\}+\frac{L}{2gc}\frac{\partial \Delta ^{2}}{\partial H} 
\end{equation}  
Now: 
\begin{equation}  
\frac{\partial ({\cal G}^{-1})^{\alpha \gamma }}{\partial H}=-\frac{\delta
^{\alpha \gamma }}{gc}H(1-\delta ^{\alpha 3})+\frac{\Omega _{n}}{gc}
\epsilon^{\alpha \gamma 3}+\frac{\delta ^{\alpha \gamma }}{2gc}\frac{\partial \Delta
^{2}}{\partial H}
\end{equation}  
It turns out therefore that the overall coefficient multiplying $\frac{%
\partial \Delta ^{2}}{\partial H}$ vanishes identically at the saddle point.
The average magnetization per site: $m\equiv M/L$ is then given by: 
\begin{eqnarray}
m&=&-\frac{2H}{\beta Lgc}\sum\limits_{q,n}
\frac{\Omega _{n}^{2}/gc-A}{A^{2}+C^{2}} \nonumber\\
&=&-\frac{1}{\beta L}\frac{H}{(gc)^{2}}\sum\limits_{q,n}\frac{\Omega
_{n}^{2}-c^{2}(q^{2}+\xi ^{-2})}{A^{2}+C^{2}} \nonumber\\
&=&-\frac{4H}{L}\sum\limits_{q}\frac{1}{\beta }\sum\limits_{n}\frac{\Omega
_{n}^{2}+z_{+}z_{-}}{(\Omega _{n}^{2}-z_{+}^{2})(\Omega _{n}^{2}-z_{-}^{2})}
\end{eqnarray}
where $z_{\pm }=i\epsilon _{\pm }(q)$ and we have used 
$c^{2}(q^{2}+\xi^{-2})=-z_{+}z_{-}$. 
The frequency sum is easily evaluated as:
\begin{eqnarray}
&&\frac{1}{\beta }\sum\limits_{n}\frac{\Omega _{n}^{2}+z_{+}z_{-}}{(\Omega
_{n}^{2}-z_{+}^{2})(\Omega _{n}^{2}-z_{-}^{2})} =\nonumber\\
&&\qquad-\frac{1}{2H}\left\{\coth \left\{\frac{1}{2}\beta \left[
\sqrt{c^{2}q^{2}+\Delta ^{2}}-H\right]\right\} \right.\nonumber\\
&&\qquad-\coth\left. \left\{\frac{1}{2}\beta \left[ 
\sqrt{c^{2}q^{2}+\Delta ^{2}}+H\right]\right\}\right\}
\end{eqnarray}

Hence we obtain the cutoff-independent result:
\begin{eqnarray}
m &=&\frac{2}{L}\sum\limits_{q}\left\{\coth \left\{\frac{1}{2}\beta
\left[ \sqrt{c^{2}q^{2}+\Delta ^{2}}-H\right]\right\} \right.\nonumber\\
&&-\coth\left. \left\{\frac{1}{2}\beta \left[ \sqrt{c^{2}q^{2}+
\Delta ^{2}}+H\right]\right\}\right\}
\label{mag}
\end{eqnarray}
which has to be considered together with Eq.(\ref{saddle}) and 
Eqs.(\ref{sp1})-(\ref{sp2}) 
that determine $\xi$ and hence $\Delta$ as a function of field and
temperature.
As a function of $H$, $m$ is odd, nondecreasing and vanishing linearly for 
$H\to 0$ as long as $T>0$. However, the two terms in curly brackets
cancel exponentially against each other when $\beta \to \infty ,$ and we
obtain the result that, as long as: $H\leq H_{c_{1}}$ holds, i.e. in the
whole Haldane phase $0\leq H\leq H_{c_{1}}$ and at 
$T=0$ the magnetization vanishes identically, as it should. 
This completes the discussion of the mean-field theory for the 
O(3) NL$\sigma$M in the Haldane phase.

\section{Discussion and Conclusions}

In Sect. II we have presented numerical results for finite chains. Our results
show in a nonambiguous way the persistence of a Haldane (gapped) phase up to
a lower critical field that can be identified with the spin gap as well as
critical behaviour, with algebraic decay of correlations, for higher fields
up to a saturation field. The shape of the magnetization curve and the
values obtained there for the spin gap and the critical exponent above the
lower critical field are in good agreement with previous work, both
numerical \ci{ST1,WH,GJL,Moreo} and analytical \ci{A91}.

In Sect. III we have developed an effective continuum field theory for the
Haldane phase. The results obtained there can be commented as follows:

i) At $T=0$ we obtain a spin gap and a vanishing magnetization up to the
lower critical field $H_{c_{1}}$. At the mean-field level the elementary
excitation spectrum consists in a triplet of magnons, and one magnon branch
goes soft at $H=H_{c_{1}}$, thus signalling an instability and the
transition to a gapless regime. Parenthetically, in a recent paper Loss and
Normand \ci{LN} have developed an approach quite similar to ours. They claim
however that the magnetization does not vanish in the spin-gap phase at the
mean-field level. On the light of the results of Sect. III such a claim seems
to be incorrect, and a completely consistent picture for the Haldane phase
appears to emerge already at the simplest mean-field level.
In its present form and with an appropriate redefinition of the coupling 
constant and of the spin-wave velocity \ci{EMPR}, the theory could apply 
also to the description of spin ladders in an external field \ci{LN} in 
the regime in which the correlation lenght is much larger than the width 
of the ladder

ii) It is interesting to remark that no singularities can develop at
finite temperatures. 
Indeed, as already remarked immediately after Eq.(\ref{xic}), setting 
$\xi ^{-2}=0$ the last term in (\ref{sp2}) would develop a severe infrared
divergence for any $\beta <\infty$ (strictly) which is altogether absent at 
$T=0.$ Therefore at $T>0$ $\xi$ remains finite and the
system is in a disordered (gapped) phase, for all values of the applied
field, as it should. To be more precise, finiteness of 
$\xi$ implies $\Delta >H$ (strictly) for all values of $H$, 
with the following consequences: a) no magnon mode will go soft at any 
$T>0$ as, in that case (see Eqs.(\ref{eps}) and (\ref{delta})), 
the spectrum of elementary excitations will be given by:
$\epsilon =\epsilon_0, \epsilon_{\pm}$ with: 
$\epsilon_{\pm} = \sqrt{\Delta^2+c^2 q^2}\pm H$
and hence: $\epsilon_{-}(q=0)=\Delta-H > 0$ 
(although exponentially small), and: b)the magnetization (see Eq.(\ref{mag}))
remains smooth all the way up to saturation, although exponentially small at
low temperatures.

What we have presented in Sect. III is a field-theoretic approach possibly at
its simplest level. At least for fields $H\lesssim H_{c_{1}}$, however, the
role of quantum fluctuations cannot be underestimated and should be taken
into account. On this direction, as well as in that of obtaining a 
consistent field-theoretic description of the gapless phase, 
work is presently in progress and we hope to report on it in 
the near future.


\begin{thebibliography}{99}

\bibitem{AMS} 
For \ comprehensive reviews, see, e.g.: i) I.Affleck, in:
\textit{Fields, Strings and Critical Phenomena}, E. Brezin and 
J. Zinn-Justin Eds., North-Holland, Amsterdam, 1990; 
ii) E. Manousakis, Rev. Mod. Phys. \textbf{63}, 1 (1991); 
iii) S.Sachdev, in: \textit{Low Dimensional Quantum
Field Theories for Condensed Matter Physicists}, S. Lundqvist, 
G. Morandi and Yu Lu Eds., World Scientific, Singapore, 1994.

\bibitem{BEMS}  A.P. Balachandran, E. Ercolessi, G. Morandi, A.M. Srivastava:
\textit{Hubbard Model and Anyon Superconductivity.} Lect. Notes in Physics
Vol.38. World Scientific, 1990.

\bibitem{Ha}  F.D.M. Haldane, Phys. Rev. Lett. \textbf{50}, 1153 (1983).

\bibitem{Ra}  R. Rajaraman: \textit{Solitons and Instantons.} 
North-Holland, 1987.

\bibitem{Au}  A. Auerbach: \textit{Interacting Electrons and Quantum
Magnetism.} Springer-Verlag, 1994.

\bibitem{Fra} E. Fradkin: \textit{Field Theories of Condensed Matter 
Systems.} Addison-Wesley, 1991.

\bibitem{Mo}  G. Morandi: \textit{The Role of Topology in Classical and 
Quantum Physics.} Springer, 1992.

\bibitem{Za2}  A.B. Zamolodchikov and A.B. Zamolodchikov, Ann. Phys. (N.Y.)
\textbf{120}, 253 (1979).

\bibitem{SR} R. Shankar and N. Read, Nucl. Phys. \textbf{B336}, 457 (1990).

\bibitem{Be}  H. Bethe, Z.Phys. \textbf{71}, 205 (1931).

\bibitem{Ma}  D.C. Mattis: \textit{The Theory of Magnetism I}, 
Springer-Verlag, 1988.

\bibitem{LSM}  E. Lieb, T.D. Schultz, D.C. Mattis, Ann.Phys. 
\textbf{16}, 407 (1961).

\bibitem{RZR}  L.P. Regnault, I. Zaliznyak, J.P. Renard and C. Vettier, 
Phys.Rev. \textbf{B50}, 9174 (1994).

\bibitem{AKLT}  I. Affleck, T. Kennedy, E. Lieb and H. Tasaki,
Comm. Math. Phys.\textbf{115}, 477 (1988).

\bibitem{Yu} J. Lou, X. Dai, S. Qin, Z. Su and Lu Yu, Cond-Mat/9904035.

\bibitem{Gr}  R.B. Griffiths, Phys. Rev.\textbf{133A}, 768 (1964).

\bibitem{De}  D.C. Dender et al., Phys. Rev. Letters \textbf{79}, 1750 (1997).

\bibitem{OA}  M. Oshikawa, I. Affleck, Phys. Rev. Lett.
\textbf{79}, 2883 (1997).

\bibitem{PB}  J.B. Parkinson, J.C. Bonner, Phys.Rev. \textbf{B32}, 4703 (1985). 

\bibitem{A91}  I. Affleck, Phys.Rev. \textbf{B43}, 3215 (1991). 

\bibitem{OYA}  M. Oshikawa, M. Yamanaka, I. Affleck, Phys. Rev. Lett.
\textbf{78}, 1984 (1997).

\bibitem{Schu1}  H.J. Schulz, Phys.Rev. \textbf{B34}, 6372 (1986).

\bibitem{A88}  I. Affleck, Phys.Rev. \textbf{B37}, 5186 (1988). 

\bibitem{CHP}  D.C. Cabra, A. Honecker and P. Pujol, Phys. Rev. Lett. 
\textbf{79}, 5126 (1997) and: Cond-Mat/9802035, Feb., 1998.

\bibitem{Ta}  H. Tasaki, quoted as ''Private Communication'' in \ci{OYA}.

\bibitem{ST2}  T. Sakai, M. Takahashi, Cond-Mat/9710327.

\bibitem{Chab}  G. Chaboussant et al., Phys. Rev. \textbf{B55}, 3046 (1997).

\bibitem{ST1}  T. Sakai, M. Takahashi, Phys.Rev. \textbf{B43}, 13383 (1991). 

\bibitem{WH} S.R. White, D.A. Huse, Phys. Rev. \textbf{B48}, 3844 (1993) 

\bibitem{WN} S.R. White, R.M. Noack, Phys. Rev. Lett. \textbf{69}, 
2863 (1992).

\bibitem{WW} S.R. White, Phys. Rev. \textbf{B48}, 10345, 1993. 

\bibitem{Spa} M.A. Martin-Delgado and G. Sierra in 
``Density Matrix Renormalization'', eds. I. Peschel et al. 
Springer-Verlag, 1999.

\bibitem{GJL} O. Golinelli, Th. Jolicoeur, R. Lacaze, 
Phys. Rev. \textbf{B50}, 3037, 1994. 

\bibitem{Grad} I.A. Gradshteyn and I.M. Ryzhik 
\textit{Tables of Integrals, Series and Products}, Academic Press, 1994. 

\bibitem{Moreo} A. Moreo, Phys.Rev. \textbf{B35}, 8562 (1987). 

\bibitem{Mi} H.J. Mikeska, J.Phys. {\bf C13}, 2913 (1980).
A.F. Andreev, V.I. Marchenko, Sov. Phys Usp. {\bf 23}, 21 (1980)

\bibitem{LN} D.Loss, B.Normand, Cond-Mat/9902104.

\bibitem{AL} S.Allen, D.Loss, Physica {\bf A239,} 47 (1997). 

\bibitem{SA} E.S. Sorensen, I. Affleck, Phys. Rev. \textbf{B49}, 15771 (1994). 
See also: I. Affleck,R.A. Weston, Phys.Rev.\textbf{B45}, 4667 (1992); 
I. Affleck, G.F. Wellmann, Phys. Rev. \textbf{B46}, 8934 (1992) and: 
E.S. Sorensen, I. Affleck, Phys. Rev. \textbf{B49}, 13235 (1994).

\bibitem{EMPR}  S. Dell'Aringa, E. Ercolessi, G. Morandi, P. Pieri, 
M. Roncaglia, Phys. Rev. Lett. \textbf{78}, 2457 (1997).


\end{thebibliography}
\end{document}